\documentstyle[prc,aps]{revtex}

\begin{document}
\draft
\title{On the Momentum Dependence\\ of the Nucleon - Nucleus Optical
Potential}
\author{M.~Kleinmann, R.~Fritz , H.\ M\"{u}ther}
\address{Institut f\"{u}r Theoretische Physik,\\ Universit\"{a}t
T\"{u}bingen,\\D-72076 T\"{u}bingen, Germany}
\author{and A.~Ramos}
\address{Departament d'Estructura i Constituents de la Materia,\\
Universitat de Barcelona, \\E-08028 Barcelona, Spain}

\date{\today}
\maketitle

\begin{abstract}
The momentum dependence of the mean-field contribution to the real part
of the optical model potential is investigated employing realistic
nucleon-nucleon interactions. Within a
non-relativistic approach a momentum dependence originates from the
non-locality of the Fock exchange term. Deducing the real part of the
optical model from a relativistic Dirac Brueckner Hartree Fock
approximation  for the self-energy of the nucleons yields an additional
momentum dependence originating from the non-relativistic reduction of
the self-energy. It
is demonstrated that large Fock terms are required in the
non-relativistic approach to simulate these relativistic features.
A comparison is made between a local density approximation for the
optical model and a direct evaluation in finite nuclei.
\end{abstract}


\section{Introduction}

The nucleon - nucleus optical potential is a very important
tool for the analysis of nuclear reaction data. Therefore a lot of
empirical information on nucleon - nucleus scattering for a
large spectrum of energies and various nuclei has been collected
and the global fits to these optical potentials \cite{pb62,hod84,ham90}
provide a very condensed information on nuclear structure properties. It is
obvious that theoreticians have devoted much interest to the study of
this important quantity. Rather than trying to give a short overview on
such investigations we refer to some recent review articles on this
subject \cite{ray92,mah91,ger85,sin75}.

{}From a theoretical point of view the real part of the optical potential
can be split into a mean-field
contribution and a dispersive correction\cite{mah85}. The name
dispersive contribution originates from the fact that this real part is
related to the imaginary part of the optical potential by a dispersion
relation. This dispersion relation yields an explicit energy-dependence
also for the real part. In principle, the dispersive contribution
could be determined from the empirical data on the imaginary component.
This would require the knowledge of the imaginary part at all energies.
For large energies, however, the imaginary part gets contributions from
nucleonic excitations, due to the strong short-range components of a
realistic NN interaction\cite{bor92},
and the excitation of sub-nucleonic degrees
of freedom, like particle production, which is beyond the scope of the
present discussion. Therefore applications of such an dispersive
optical-model analysis typically consider only the imaginary part at
low energies \cite{tor90} and use some kind of subtracted dispersion
relation\cite{mah85}.

If one tries to describe the remaining mean-field contribution of the
optical potential in terms of a local potential to be used in a
Schroedinger equation, one observes a significant dependence of this
mean-field part on the energy or asymptotic momentum of the scattered
nucleon\cite{bor92,mas91}. There are two possible sources for this
momentum dependence. Within a non-relativistic approach one may
attribute this momentum dependence to the non-locality of the exchange
part in the Hartree-Fock (HF) or Brueckner-Hartree-Fock (BHF)
approximation for the mean-field part of the optical potential.
Transforming this non-local HF potential to an equivalent local
potential leads to such a momentum dependence \cite{sin75,ps64} (see also
chapter 2).

Another source for the momentum-dependence of the mean-field part of
the optical potential is of relativistic origin. Relativistic
approaches have been very successful to describe bulk properties of
nuclei. This is true for more phenomenological models, like the $\sigma
- \omega$ model of Walecka and Serot \cite{serot}, but also for
microscopic Dirac-BHF calculations based on realistic nucleon-nucleon
(NN) forces\cite{rupr,shakin,BM84,malf}. Characteristic for all these
calculations is the relativistic structure of the nucleon self-energy,
which contains a large Lorentz-scalar and a large time-like vector
component. In a simple Dirac-Hartree calculation these components are
local and energy independent. Reducing the two coupled Dirac equations
for the upper and lower component to one Schroedinger equation, one
derives from these scalar and vector components a central Schroedinger
equivalent potential (SEP), which depends on the energy or momentum of
the nucleon \cite{jam80}.

The scalar and vector components of the relativistic self-energy
obtained in a Dirac-HF or Dirac-BHF calculation are non-local.
Therefore both sources for the momentum dependence of the mean-field part
of the optical model discussed above get relevant. It is one aim of the
present investigation to explore the importance of these two effects
and their interplay. This study is based on the self-consistent
Dirac-BHF calculations, which were recently performed directly for
finite nuclei\cite{fri1,fri2}.

There have been many very successful attempts to deduce the mean field
part of the optical potential directly from the NN T-matrix or G-matrix
\cite{ray92} within a non-relativistic \cite{ri83,co90,ar90,el90} and
relativistic framework \cite{ho85,co90b,ot91}. The impulse
approximation which is used in most of these investigations is rather
successful in particular for proton scattering at energies above 100
MeV. Medium effects are typically derived from nuclear matter and taken
into account in a local density approximation \cite{li93}.

The main emphasis of the present investigation is to study the
differences and similarities in non-relativistic and relativistic
approaches in particular at lower energies. These energies are
important for the extrapolation of the single-particle potential at
negative energies, connecting the optical potential to the binding
properties. For that purpose we consider simple parameterizations of
G-matrix approaches, derived from realistic NN forces. In the case of
the non-relativistic approach this will be the so-called M3Y
parameterization\cite{m3y}. In the case of the relativistic treatment this
will be the effective meson parameterization of \cite{fri1,fri2}. This
enables us to study and compare the origin of the momentum dependence
obtained in the non-relativistic and relativistic approach in detail.
Furthermore we would like to compare the results for nucleon-nucleus
scattering obtained from a non-local Dirac self-energy which is
calculated directly for the finite nuclei to results which are obtained
in the frequently used local-density approximation.

After this introduction we will briefly recall the connection between
the non-locality and the momentum dependence of an equivalent local
mean field in section 2. In section 3 we will shortly describe the
momentum dependence of the SEP derived from a local Dirac self-energy.
There we will also describe an efficient method to calculate scattering
phase shifts from a non-local relativistic self-energy. The results of
our investigation are presented and discussed in section 4 and the
conclusions are summarized in a final section 5.

\section{Non-Locality and Exchange Effects}

More than 15 years ago Bertsch et.al.~determined a very efficient
parameterization of an effective NN interaction from the G-matrix
evaluated for realistic forces \cite{m3y}. These so-called M3Y
potentials are given in terms of local Yukawa potentials. The central
part of the interaction can be written as a function of the distance
$r$ between the interacting nucleons
\begin{equation}
\hat V_{\hbox{M3Y}} = V_{0}(r) + V_{\tau}(r)
\vec{\tau_{1}} \cdot \vec{\tau_{2}} + V_{\sigma}(r)
\vec{\sigma_{1}} \cdot \vec{\sigma_{2}} + V_{\sigma\tau}
\vec{\sigma_{1}} \cdot
\vec{\sigma_{2}} \vec{\tau_{1}} \cdot \vec{\tau_{2}}\; . \label{eq:2.1}
\end{equation}
The operators $\vec{\tau_{i}}$ and $\vec{\sigma_{i}}$ denote the Pauli
matrices acting on the isospin and spin of nucleon $i$, respectively.
The radial shapes of the functions $V_{\alpha}(r)$ are given in terms
of 2 Yukawa potentials
\begin{equation}
V_{\alpha}(r) = \sum_{k=1}^2 C_{\alpha}^{(k)} \frac{\exp{(-\mu_{k}r)}}
{\mu_{k}r} \; ,\label{eq:2.2}
\end{equation}
only for the spin-isospin channel $V_{\sigma\tau}$ a third Yukawa term
has been added, which represents the central component of the
pion-exchange potential
\begin{eqnarray}
\Delta V_{\sigma\tau} & = & \frac{1}{3} \frac{g_{\pi}^2}{4\pi}
\frac{m_{\pi}^2}{4m^2} m_{\pi} \frac{\exp{(-\mu_{\pi}r)}}{\mu_{\pi}r}
\nonumber \\
& = & C_{\pi}\frac{\exp{(-\mu_{\pi}r)}}{\mu_{\pi}r} \; .\label{eq:2.3}
\end{eqnarray}
Here $\mu_{\pi} = m_{\pi}/(\hbar c) = 0.7072$ fm$^{-1}$ and $C_{\pi} =
3.4877$ MeV, which corresponds to a pseudoscalar coupling constant for
the pion of $g_{\pi}^2 / (4\pi) \approx 13.9$. Assuming for the mass
parameter $\mu_{k}$ in Eq.(\ref{eq:2.2}) values of 4 fm$^{-1}$ and 2.5
fm$^{-1}$, for $k=1$ and 2, respectively, the 8 strength parameter
$C_{\alpha}^{(k)}$ were adjusted to reproduce G-matrix elements of the
Reid potential \cite{reid} and the Elliot potential \cite{ellio}
in an oscillator basis. The resulting parameters are listed in table 1.

For the mean-field contribution of the optical potential for scattering
of nucleons on closed shell nuclei with N=Z only the spin- and isospin
averaged part of the NN interaction is relevant. For the direct
interaction leading to the Hartree part of the potential this is
identical to the scalar isoscalar term in Eq.(\ref{eq:2.1})
\begin{eqnarray}
V_{\hbox{dir}} & = & V_{0} (r) \nonumber \\
& = & 7999.0 \frac{\exp{(-4r)}} {4r} - 2134.25 \frac{\exp{(-2.5r)}}
{2.5r}\; .\label{eq:2.4}
\end{eqnarray}
The resulting direct contribution to the optical potential is given by
\begin{equation}
U_{\hbox{dir}}(r) = \int d^3r' \rho_{b}(r') V_{\hbox{dir}} (\vert \vec r -
\vec r' \vert) \label{eq:2.5}
\end{equation}
with the nuclear density $\rho_{b}(r) = \rho (r,r)$
obtained in the Hartree-Fock approximation
\begin{equation}
\rho (r,r') = \sum_{i=1}^N \Phi_{i}^*(r) \Phi_{i}(r') \; .
\label{eq:2.6}
\end{equation}
Here $\Phi_{i}$ stands for the single-particle wavefunctions occupied
in the Hartree-Fock approximation for the target nucleus.
The exchange part of the NN interaction is evaluated as
the scalar isoscalar part of the Fierz transformation \cite{czers} of
Eq.(\ref{eq:2.1}) and is given as
\begin{eqnarray}
V_{\hbox{exc}} & = & -\frac{1}{4} \left[ V_{0}(r) + 3 V_{\tau}(r) + 3
V_{\sigma}(r) + 9 V_{\sigma\tau}(r) \right] \nonumber \\
& = & 4631.375 \frac{\exp{(-4r)}} {4r} - 1787.125 \frac{\exp{(-2.5r)}}
{2.5r} - 7.8474 \frac{\exp{(-0.7072r)}} {0.7072r} \; .\label{eq:2.7}
\end{eqnarray}
This local parameterization of the G-matrix and slight modifications of
it have been used for various nuclear studies including the evaluation of
the optical potential for heavy-ion scattering \cite{khoa}. The
exchange part of the NN interaction yields the Fock contribution to the
mean field which is non-local
\begin{equation}
U_{\hbox{exc}}( r, r') = \rho (r,r') V_{\hbox{exc}} (\vert \vec r -
\vec r' \vert) \label{eq:2.8}
\end{equation}
The local equivalent potential is defined by the requirement
\begin{equation}
U_{\hbox{exc}}^{loc} (r) \Psi_{E}(r) = \int d^3r' U_{\hbox{exc}}( r, r')
\Psi_{E}(r') \; , \label{eq:2.9}
\end{equation}
where $\Psi_{E}(r)$ is the solution of the Schroedinger equation for a
nucleon at energy $E$ moving in the single-particle potential
$U=U_{\hbox{dir}}+U_{\hbox{exc}}$. Negele and Vautherin \cite{nege}
demonstrated that Eq.(\ref{eq:2.9}) can well be approximated by
\begin{equation}
U_{\hbox{exc}}^{loc} (r)  = \int d^3r' V_{\hbox{exc}} (\vert \vec r -
\vec r' \vert) \rho( r, r') \frac {\sin{k(r)\vert \vec r -\vec r'
\vert} } {k(r)\vert \vec r -\vec r'\vert} \; , \label{eq:2.10}
\end{equation}
with the local approximation to evaluate $k$
\begin{equation}
k^2(r) = \frac {2m}{\hbar^2} \left[ E - U(r) \right] \; .
\label{eq:2.11}
\end{equation}
It is evident from Eq.(\ref{eq:2.9}) that the local equivalent
potential depends on the energy $E$ or the momentum $k(\infty)$
of the scattered nucleon. In
Eq.(\ref{eq:2.10}) this momentum-dependence is hidden in the local
momentum $k(r)$, which must be calculated in a self-consistent way from
Eq.(\ref{eq:2.11}).

The mixed density $\rho(r,r')$ entering the exchange part of the mean
field potential is frequently approximated by the so-called Slater
approximation obtained in nuclear matter
\begin{equation}
\rho(r,r') = \rho_{b}\left( \frac{\vec r + \vec r'}{2}\right)  \frac {3}
{\vert \vec r -\vec r'\vert k_{F}} j_{1}\left(\vert \vec r -\vec r'
\vert k_{F}\right) \; , \label{eq:2.12}
\end{equation}
with the Spherical Bessel function $j_{1}(x)$ \cite{abramo}
and the local Fermi
momentum, which is related to the baryon density $\rho_{b}$ at the central
point between $\vec r$ and $\vec r'$ by the nuclear matter relation
\begin{equation}
k_{F}^3 = \frac{3 \pi^2}{2} \rho_{b} \label{eq:2.13}
\end{equation}

\section{Dirac Equation and Schroedinger Equivalent Potential}

A simple parameterization of Dirac-Brueckner-Hartree-Fock (DBHF)
calculations for nuclear matter employing realistic NN interactions has
been presented in Ref.\cite{fri1,fri2}. For each density $\rho_{b}$ two
coupling constants for a scalar ($G_{\sigma}$) and vector meson
($G_{\omega}$)
exchange have been determined such that a simple Dirac Hartree
calculation within the $\sigma - \omega$ model reproduces the bulk
properties of nuclear matter at this density obtained from DBHF. The
density dependence of the coupling constants reflects the
density dependence of the correlations taken into account in DBHF.

Using these density dependent coupling constants one can determine the
nucleon self-energy in a Dirac-Hartree (DH) approximation
\begin{equation}
\hat\Sigma_{\hbox{DH}} = \Sigma^s_{\hbox{DH}} - \gamma^0
\Sigma^0_{\hbox{DH}} \label{eq:3.1}
\end{equation}
with a scalar and vector contribution given by
\begin{eqnarray}
\Sigma^s_{\hbox{DH}} (r) & = & - G_{\sigma}(r)
m_\sigma \int_0^{\infty} r'^2 dr' G_{\sigma}(r')\rho_s(r')
	    \widetilde{I}_0 (m_\sigma r_<) \widetilde{K}_0
         (m_\sigma r_>)  \label{eq:hart1}\\
 &  &  \nonumber \\
 \Sigma^0_{\hbox{DH}} (r) & = & -G_{\omega}(r)
  m_\omega \int_0^{\infty} r'^2 dr' G_{\omega}(r')\rho_{b} (r')
	    \widetilde{I}_0 (m_\omega r_<) \widetilde{K}_0 (m_\omega r_>)
         \label{eq:hart2}
\end{eqnarray}
In these equations the coupling constants $G_{\sigma}(r)$ and
$G_{\omega}(r)$ denote the density dependent coupling constants discussed
above calculated at the nuclear density which is equal to the baryon
density $\rho_{b}$ of the nucleus at the position $r$. The meson mass
parameters were chosen to be $m_{\sigma}$=550 MeV and $m_{\omega}$=783
MeV in accordance with the realistic OBE potentials\cite{rupr}.
The functions $\widetilde{I}_L (x)$ and $\widetilde{K}_L (x)$ arise
from the multipole expansion of the meson propagator in coordinate space
and are defined using the modified spherical Bessel functions $I$ and
$K$\cite{abramo}:
\begin{equation}
    \widetilde{I}_L (x) = \frac{I_{L+1/2} (x)}{\sqrt{x}} \; \; \: \:
    \widetilde{K}_L (x) = \frac{K_{L+1/2} (x)}{\sqrt{x}} \; .
\end{equation}
The baryon density $\rho_{b}$ and the scalar density $\rho_{s}$ are
defined by
\begin{eqnarray}
    \rho (r)& = &\frac{1}{4 \pi} \sum_a (2{\jmath}_a+1)
		\left[ g_a^2 (r)  + f_a (r)^2  \right] \nonumber\\
    \rho_s(r)& = &\frac{1}{4 \pi} \sum_a (2{\jmath}_a+1)
		\left[ g_a^2 (r)  - f_a (r)^2  \right] \label{eq:dens}\; ,
\end{eqnarray}
with a summation index $a$ ranging over all occupied shells. The
functions $g_{a}$ and $f_{a}$ refer to the small and large components
of the Dirac spinor
\begin{equation}
   <r|\alpha> =  \Psi_{\alpha} (\vec{r})  =
      \left( \begin{array}{c}
	      g_a(r)  \\
	      -i f_a(r) \; \vec{\sigma} \cdot \hat{r}
	      \end{array}   \right)
      {\cal Y}_{\kappa_a, m_a} (\Omega) \; , \label{eq:spinor}
\end{equation}
using the notation of Ref.\cite{fri2}. The Dirac spinors are obtained
by solving the Dirac equation including the nucleon self-energy defined
in Eq.(\ref{eq:3.1}). The definition of the self-energy and the
solution of the Dirac equation are connected by a self-consistent
requirement. Note, however, that in the present model, the baryon
density $\rho_{b}$ obtained from solving the Dirac equation enters the
definition of the self-energy not only as displayed explicitly in
Eqs.(\ref{eq:hart1}) and (\ref{eq:hart2}) but also via the density
dependent coupling constants.

The contributions to the self-energy $\Sigma^s_{\hbox{DH}}$ and
$\Sigma^0_{\hbox{DH}}$ are local in the Dirac Hartree approximation.
The Dirac equation corresponds to two coupled equations for the large
($g$) and small ($f$) component of the resulting Dirac spinor. One can
eliminate the small component in a standard way and obtain a
Schroedinger equation for a wavefunction which at large distances
agrees with the large component of the Dirac spinor \cite{jam80,jam89}.
The central part of the potential in this Schroedinger equation, called
the central part of the Schroedinger Equivalent Potential (SEP) can be
written
\begin{equation}
U_{SEP}(r) = \Sigma^s(r) - \frac{\tilde E}{m}\Sigma^0(r) +
\frac{\left(\Sigma^s(r)\right)^2 - \left(\Sigma^0(r)\right)^2 +
U_{\hbox{Darwin}}(r)}{2m} \label{eq:sep}
\end{equation}
with
\begin{eqnarray}
U_{\hbox{Darwin}}(r) & = & \frac{3}{4} \left[
\frac{1}{D(r)}\frac{d\,D(r)}{dr}\right]^2 - \frac{1}{rD(r)}\frac{d\,D(r)}{dr}
-\frac{1}{2D(r)} \frac{d^2D(r)}{dr^2}\; , \nonumber\\
D(r) & = & m + E + \Sigma^s(r) + \Sigma^0(r) \; .
\end{eqnarray}
Therefore we see that this SEP depends on the energy of the scattered
nucleon $E = \tilde E -m$ or its asymptotic momentum.
This is true already if we
employ the Dirac - Hartree approximation, i.e. deal with local
functions for the scalar and vector component of the nucleon
self-energy.

If the self-energy is evaluated in the Dirac-Hartree-Fock approximation,
the contributions are non-local. In order to evaluate the scattering
observables obtained from a solution of the Dirac equation with such a
non-local self-energy, we use the technique already described in
Ref.\cite{fri2}. In this method the radial functions $g_{a}(r)$ and
$f_{a}(r)$, which define the stationary solutions of the Dirac
equation (see Eq.(\ref{eq:spinor})) are expanded in a discrete basis of
spherical Bessel functions. The wave numbers for this basis are chosen
such that this discrete basis is a complete orthonormal basis in a sphere
of radius $D$. This radius is chosen to be large enough that the results
for the wavefunctions obtained at positive energies show the asymptotic
form at $r\approx D$. A typical value for $D$ is $D=20$ fm for the light
nuclei being considered here. With this expansion for the wavefunctions
the Dirac equation is rewritten in form of an
eigenvalue problem and the eigenvalues ($E_{a}$) and eigenvectors (the
expansion coefficients for $g_{a}$ and $f_{a}$) are determined by a
simple matrix diagonalisation \cite{thesis}. The expressions for the
matrixelements of this matrix to be diagonalized are published
elsewhere \cite{fri2,thesis}.

Due to the choice of the basis for the expansion (Bessel functions,
which vanish at $r=D$: $j_{l}(k_{n}D) =0$) also the wavefunctions
resulting from this diagonalisation must vanish at $r=D$. This implies
that
we obtain results for the bound single-particle states and for those
selected energies of scattering states $E_{\nu}$ for which the
wavefunction vanishes at $r=D$. Assuming that the wavefunction has
reached its asymptotic form, see discussion above, we can deduce the
phase shifts $\delta_{l}$ at these selected energies from the relation
\begin{equation}
g_{\nu}(D) \; = \; \cos{\delta_{l}}\, j_{l}(k_{\nu}D) -
\sin{\delta_{l}}\, y_{l}(k_{\nu}D) \; = \; 0 \; ,
\label{eq:phase}
\end{equation}
where $y_{l}$ stands for the Neumann function and
the asymptotic wavenumber $k_{\nu}$ is related to the energy
$\tilde E_{\nu}$ by
\begin{equation}
\tilde E_{\nu} \; = \; E_{\nu} + m \; = \; \sqrt{k_{\nu}^2 + m^2} \; .
\label{eq:enek}
\end{equation}
This method for the calculation of scattering phase shifts is a
relativistic extension of the scheme employed in
Refs.\cite{boro92,gua82} for non-local potentials in a Schroedinger
equation.

Also in the case of the Dirac-Hartree-Fock approximation one can try
to find a local potential which used in a Schroedinger equation yields
the same result for the scattering phase shifts as obtained from the
Dirac-Hartree-Fock approximation. It is evident that this SEP will
exhibit an energy dependence due to the Dirac effects but also due to
the non-local Fock exchange terms as discussed in the previous section.

\section{Results and Discussion}

A parameterization of the Dirac-Brueckner-Hartree-Fock G-matrix
calculated in nuclear matter for the version $A$ of the Bonn OBE
potential \cite{rupr} has been presented in Ref.~\cite{fri2}. Employing
this parameterization of the G-matrix in terms of a density-dependent
$\sigma$ and $\omega$ meson exchange in Dirac-Hartree calculation of
finite nuclei, results were obtained for the binding energy and radius
of closed shell nuclei like $^{16}O$ and $^{40}Ca$, which were in fair
agreement with the experimental data \cite{fri1,rolf}. The baryon
densities and scalar densities obtained in such calculations for $^{16}O$
and $^{40}Ca$ are displayed in figures 1 and 2, respectively. These density
distributions will be the starting point of our further investigations.

As a first approach we consider the scalar $\Sigma^s_{\hbox{DH}}$ and
vector $\Sigma^0_{\hbox{DH}}$ contribution to the nucleon self-energy
in the Dirac-Hartree approximation, which are related to these
densities via the relations Eqs.(\ref{eq:hart1}) and (\ref{eq:hart2}).
{}From these local scalar and vector contributions one can calculate a
local SEP as discussed in the preceding section (see
Eq.(\ref{eq:sep})). These local SEP are displayed in the lower parts of
figures 1 and 2 for two different energies of the scattered nucleon. The
attraction of the SEP for nucleons with an energy of 100 MeV is reduced
by more than a factor 2 as compared to the SEP for nucleons at an
energy of 5 MeV.

This energy dependence of the SEP, which is of course well known, is
very important to obtain predictions for the elastic scattering which
are in agreement with the experimental data. As an example we show in
the upper parts of figures 3 and 4 results for the $l=0$ phase shifts
for elastic scattering of nucleons at various energies on $^{16}O$
and $^{40}Ca$. As a bench-mark, representing the experimental data, we
also show in these figures the phase-shifts obtained from a fit to
the experimental data \cite{ham90}. An acceptable agreement with these
``experimental'' values for the phase shifts is obtained, if we
evaluate the phase shifts in the Dirac-Hartree approximation or,
equivalently, consider the SEP for each energy consistently. If,
however, the SEP calculated for a nucleon energy of 5 MeV would be used
for the scattering of nucleons at higher energies as well, one obtains
a substantial deviation (see dashed curves in the upper parts of
figures 3 and 4).

As an alternative approach we now consider the non-relativistic
parameterization of the G-matrix provided by the M3Y approach
\cite{m3y}. It should be mentioned, however, that other non-relativistic
parameterizations \cite{czers} show similar features. As we want to
explore the differences between a relativistic and non-relativistic
evaluation of the mean-field contribution to the optical potential,
independent on the underlying baryon density, we use the same density
distribution as in the Dirac-Hartree approach. For these densities we
determine the scattering potential, including the exchange term in the
local form of Eq. (\ref{eq:2.10}) with the Slater approximation of
Eq.(\ref{eq:2.12}) for the mixed density $\rho(r,r')$.

The resulting local potentials are also displayed in figures 1 and 2
(dashed curves) again for nucleons with energies of 5 and 100 MeV. The
shapes of the M3Y potentials deviate from those obtained in the
Dirac-Hartree approach in a distinct manner. The M3Y potentials tend to
be less attractive at the surface and more attractive close to the
center. The energy dependence of these local potentials, however, seems
to be quite similar to the one obtained in the Dirac-Hartree approach.

This is supported by the inspection of the scattering phase shifts in
the lower parts of figures 3 and 4. Also for the M3Y potentials we
obtain an acceptable agreement with the ``experimental'' data if the
energy dependence of the local potential is taken into account. If,
however, a scattering potential determined for an energy of 5 MeV is
used for the scattering of nucleons with higher energies as well, the
calculated phase shifts are far above the experiment at energies around
100 MeV.

The results obtained for the phase shifts display a remarkable
similarity comparing the Dirac-Hartree and the M3Y approach. Similar
results are also obtained for other angular momenta and the total cross
section. For angular momenta $l$ different from zero, and consequently
also for the total cross section, the Dirac-Hartree approach tends to
give slightly larger results. This can be related to the larger radii
of the Dirac-Hartree potentials.

At this stage one may expect that a combination of the energy
dependence in the local potential due to the Dirac effects and due to
the Fock exchange effects, as considered in a Dirac-Hartree-Fock
calculation should lead to an energy dependence of the calculated phase
shifts, which is larger than in the Dirac-Hartree or in the
non-relativistic Hartree-Fock approximation and therefore exceeds the
energy dependence of the empirical data.

This, however, is not the case. To demonstrate this, we use the
Dirac-Hartree-Fock
parameterization of the G-matrix evaluated in nuclear matter again for
version $A$ of the OBEP (including $\pi$ and density dependent $\sigma$
and $\omega$ exchange). We take the same density distributions for
$^{16}O$ and $^{40}Ca$ as discussed above. (The self-consistent
Dirac-Hartree-Fock densities deviate from the Dirac-Hartree densities
used here and are not considered as we would like to study the
differences in the self-energies which are not due to different
densities.) For these ingredients one can calculate the non-local
Dirac-Hartree-Fock self-energy and evaluate the phase using the
technique discussed in section 3.

Results for the $l=0$ phase shifts are displayed in figure 5 for the
case of the nucleus $^{40}Ca$. These phase-shifts are rather close to
those obtained in the Dirac-Hartree approximation. In fact the energy
dependence obtained in the Dirac-Hartree-Fock approximation is even slightly
weaker than the one obtained in the Dirac-Hartree calculation.

Figure 5 also shows results for the scattering phase shifts calculated
in a local density approximation (LDA) directly from the nucleon
self-energy in nuclear matter \cite{li93}.
This means that one determines the different components of the
self-energy for various densities of nuclear matter
and identifies the local self-energy at position $\vec r$ with
the corresponding nuclear matter result calculated for the baryon
density at this position. For the scalar part of the self-energy this
means
\begin{equation}
\Sigma^s_{\hbox{LDA}} (r,E) = \Sigma^s_{\hbox{Nuc.Mat.}}(\rho (r),k)
\label{eq:lda}
\end{equation}
where $\Sigma^s_{\hbox{Nuc.Mat.}}(\rho ,k)$ is the scalar part of the
nucleon self-energy calculated in nuclear matter of density $\rho$ for
a nucleon with momentum $k$, which is equal to the asymptotic momentum
of the scattered nucleon. Equivalent expressions define the other
components of the self-energy. Inserting this expression into
Eq.~(\ref{eq:sep}) one obtains the SEP in this local density
approximation. The phase shifts can be calculated either directly from
the solution of the Dirac equation with a self-energy according to
Eq.~(\ref{eq:lda}) or by solving a Schroedinger equation for the SEP.

{}From figure 5 we see that the LDA yields results which are in good
agreement with the phase shifts calculated in Dirac-Hartree and
Dirac-Hartree-Fock approximation. Very similar agreement is also
obtained in the case of the nucleus $^{16}O$, which is not presented
here. In particular the energy or momentum
dependence of the SEP seems to be contained already in the nuclear
matter approximation. This can also be seen from the plot of the SEP in
figure 2. The LDA yields predictions for the local equivalent
potential, which are in very good agreement with the Dirac-Hartree
results. In particular the energy dependence is well reproduced. The
radial shapes deviate in a characteristic way. The LDA reflects the
fluctuations in the baryon density more drastically and exhibits a
smaller radius than the Dirac-Hartree approximation. This is due to the
suppression of finite range effects in the nucleon-nucleon interaction,
which is characteristic for local density approximations \cite{czers}.

{}From this comparison one can see that the discussion of nuclear matter
results using a local density approximation yields valuable
information on general features of finite nuclei, like the momentum
dependence of the optical model. It must be kept in mind, however, that
the local-density approximation in general is not very reliable in
predicting specific observables with high accuracy \cite{czers}. As an
example we would like to mention that the LDA overestimates the
calculated binding energy per nucleon by 3.2 MeV and 2.3 MeV for
$^{16}O$ and $^{40}Ca$, respectively. Furthermore it yields smaller
radii for the ground states of these nuclei than the corresponding
direct calculation for the finite nuclei (see also figure 2 and
discussion of the SEP above).

In our investigation we find that the energy- or momentum-dependence of the
optical potential can be described within a non-relativistic as well
as in a relativistic framework. This is the same situation as in the
empirical approaches to describe nucleon-nucleus scattering
\cite{ray92}. The physical origin of the
momentum-dependence of the local equivalent potentials displayed in
figures 1 and 2 is quite different in the two approaches. Within the
non-relativistic description, the momentum dependence is determined by
the strong Fock exchange term in the M3Y parameterization. If one
calculates the binding of nuclear matter with this effective
interaction (see Eq.(\ref{eq:2.1}), one finds that the dominating
contribution to the potential energy arises from the Fock exchange terms.
This is displayed in figure 6, where the direct
Hartree contribution to the potential energy of nuclear matter is
compared to the total Hartree-Fock result for various densities. One
finds that the Hartree term only yields about 25 percent of the total
attraction.

These strong exchange terms are required to provide a sufficient
momentum dependence for the optical potential. In order to identify the
most important contribution to this momentum dependence figure 7 shows
the momentum dependence of the single-particle potential in nuclear
matter including the various components of the M3Y interaction. As our
discussion of the local density approximation above demonstrated that
the main ingredients of the energy-dependence are already obtained for
nuclear matter, we will concentrate our discussion on nuclear matter. From
the analysis displayed in figure 7 it is obvious that the momentum
dependence is dominated by the components of medium range in the
exchange part of the M3Y interaction. The pion-exchange contribution is
much weaker and the repulsive short-range components increase the
attraction for larger momenta. This medium range component in
Eq.(\ref{eq:2.7}) is mainly due to $V_{\tau}$ and $V_{\sigma\tau}$.
Translated into the language of mesons this means that it is due to the
exchange of an isovector meson with a mass of around 500 MeV. As such a
meson does not exist, we must consider this as an effective meson to
parameterize the effects of correlations or other features contained in
the G-matrix.

It is very difficult to decide whether such strong exchange effects
required in the M3Y or other non-relativistic parameterizations are
realistic or just an artefact of this parameterization. The only
information available from NN scattering data and microscopic
calculations of NN interactions is the information about the sum of
direct and exchange term and therefore the separation into a direct and
an exchange term is to some extent model-dependent. There are, however,
some indications that such strong exchange effects do not show up
within the relativistic framework.

It is worth noting that in relativistic parameterizations of the
G-matrix, the self-energy is dominated by the direct
Hartree-contributions. This is valid for the Dirac-Hartree-Fock
parameterization of \cite{fri2}, which we have discussed so far. The
same is also true for the more complete parameterization of the
G-matrix derived from the Bonn potential of \cite{elsen} and for the
independent analysis of the group in Groningen \cite{malf,mal94}. The
Fock- or exchange-terms yield corrections to the Hartree terms, which
are as large as 30 percent of the Hartree term. However, one does not
meet a situation that the exchange terms are more important as it is the
case in the non-relativistic framework discussed above.

This is also reflected by the fact that the momentum or
energy-dependence of the SEP derived from the
Dirac-Brueckner-Hartree-Fock self-energy in nuclear matter is dominated
by the energy dependence displayed directly in Eq.(\ref{eq:sep}). This
is reflected in figure 8, where the momentum dependence of the SEP in
nuclear matter is displayed with (solid lines) and without (dashed
lines) taking into account the momentum dependence of the Dirac
self-energy. Both, for the Bonn potential and for the independent
analysis of the Groningen group, one observes that the momentum
dependence of the Dirac self-energy, which reflects the Fock exchange
effects, is a small correction and tends to reduce the momentum dependence
of the SEP.

\section{Conclusions}

Results for the real part of the optical potential for nucleon -
nucleus scattering at low energies (E $<$ 100 MeV) are discussed within
a non-relativistic framework and the relativistic Dirac-Hartree-Fock
approximations. The investigations are based on relativistic and
non-relativistic parameterizations of a G-matrix derived from realistic
NN interactions. Both approaches lead to local energy-dependent
potentials, which, employed in a Schroedinger equation, reproduce the
bulk features of the energy dependence of the empirical data for
nucleon scattering on $^{16}O$ and $^{40}Ca$. However, for the same
density distribution of the target nucleons different local equivalent
potentials are obtained.

A local density approximation (LDA) contains the main features of the energy
dependence of the optical potential. The LDA fails to reproduce the
precise shape of the potentials leading to smaller radii. Also it does
not provide accurate results for the binding energy.

Both kinds of approaches yield a similar energy- or momentum
dependence. Therefore, analogous to the empirical studies of nucleon
nucleus scattering, also the present investigation cannot give a
definite answer to the question whether relativistic features are
needed to describe these processes. The origin of the energy dependence
is quite different in the two approaches. Within the non-relativistic
scheme strong Fock exchange terms are required to obtain an energy
dependence of the optical model as required from the empirical data.
In the relativistic framework the energy dependence is due to the
reduction of the Dirac equation to a Schroedinger equation and Fock
exchange terms play a minor role. Therefore one may understand the
strong exchange effects of the non-relativistic approach as a tool to
incorporate the relativistic features.

It should be useful to study the optical potential for the elastic
scattering of heavy ions \cite{khoa} to investigate whether the
empirical data for these systems allow to distinguish between the
relativistic and the non-relativistic approach.

This investigation has been supported by the ``Graduiertenkolleg
Struktur und Wechselwirkung von Hadronen und Kernen'' (DFG, Mu 705).

\begin{table}
\caption{Parameterization of the different central terms of the M3Y potential
(see Eq.(1)), using the parameterization of Eq.(2). Note that for the
$\sigma\tau$ channel the pion contribution of Eq.(3) has to be added.}
\label{tab:m3y}
\begin{center}
\begin{tabular}{crrrrr}
&&&&&\\
$k$ & \multicolumn{1}{c}{$\mu_{k}$ [fm$^{-1}$]}
&\multicolumn{1}{c}{$C_{0}^{(k)}$ [MeV]}
&\multicolumn{1}{c}{$C_{\tau}^{(k)}$ [MeV]}
&\multicolumn{1}{c}{$C_{\sigma}^{(k)}$ [MeV]}
&\multicolumn{1}{c}{$C_{\sigma\tau}^{(k)}$ [MeV]}\\
&&&&&\\
\hline
&&&&&\\
1 & 4.0 & 7999.00 & -4885.50 & -2692.25 & -421.25 \\
2 & 2.5 & -2134.25 & 1175.50 & 478.75 & 480.00 \\
&&&&&\\
\end{tabular}
\end{center}
\end{table}

\clearpage
\begin{figure}[h]
\vspace{10 truecm}
\caption{Density distribution and optical potential for $^{16}O$. The
density distribution displayed in the upper part of the figure has been
obtained from the Dirac Hartree calculation of Ref.[20]. For this
distribution the local equivalent optical potential of nucleon
scattering have been calculated for energies of 5 MeV and 100 MeV,
using the Dirac Hartree (solid line) and the non-relativistic folding
potential calculated for the M3Y potential (dashed curve)}
\label{fig:fig1}
\end{figure}

\begin{figure}[h]
\vspace{10 truecm}
\caption{Density distribution and optical potential for $^{40}Ca$.
The lower part of the figure also displays the local density
approximation to the optical potential defined in Eqs.(24) and (20).
Further details see figure 1}
\label{fig:fig2}
\end{figure}
\clearpage
\begin{figure}[h]
\vspace{10 truecm}
\caption{Phase shifts for $l=0$ scattering on $^{16}O$ at various
energies of the scattered nucleon. Results of the Dirac Hartree
calculation are displayed in the upper part while those obtained from
the non-relativistic M3Y potential are given in the lower part of the
figure. Beside the results of the consistent calculations (solid lines)
also phase-shifts obtained for the equivalent local potential at a
fixed energy of 5 MeV (dashed line) are presented. For comparison also
the phase-shifts obtained from the empirical fit of Ref.[3] are
displayed as ``experimental'' data.}
\label{fig:fig3}
\end{figure}

\begin{figure}[h]
\vspace{10 truecm}
\caption{Phase shifts for $l=0$ scattering on $^{40}Ca$ at various
energies. Further details see figure 3.}
\label{fig:fig4}
\end{figure}
\clearpage
\begin{figure}[h]
\vspace{10 truecm}
\caption{Calculated $l=0$ phase shifts for scattering on $^{40}Ca$.
Results are displayed for the Dirac-Hartree, the Dirac-Hartree-Fock and
the local density approximation for the Dirac-Hartree approach.}
\label{fig:fig5}
\end{figure}

\begin{figure}[h]
\vspace{10 truecm}
\caption{Contributions to the binding energy of nuclear matter at
various densities calculated from the direct part of the M3Y
interaction (Hartree) and from the direct plus exchange part
(Hartree-Fock). The kinetic energy is not included.}
\label{fig:fig6}
\end{figure}
\clearpage
\begin{figure}[h]
\vspace{10 truecm}
\caption{Momentum dependence of the single-particle potential in
nuclear matter at saturation density ($k_{F}$ = 1.36 fm$^{-1}$)
calculated for the M3Y potential. Results are displayed including the
total exchange part of the interaction (solid line), the medium and
short-range part (dashed line) and the short-range part only (line with
circle labels). Displayed is the energy difference to the potential
energy at momentum $k=0$ in each case.}
\label{fig:fig7}
\end{figure}

\begin{figure}[h]
\vspace{10 truecm}
\caption{Momentum dependence of the Schroedinger equivalent potential (SEP)
calculated according to Eq.(20) in
nuclear matter at saturation density ($k_{F}$ = 1.36 fm$^{-1}$).
Results are displayed which account for the momentum dependence of the
scalar and vector terms in the self-energy (solid lines) and those for
which the scalar and vector terms calculated at momentum $k=k_{F}$ were
taken for all momenta (dashed lines). Results derived from the
Bonn A potential (Ref.[20]) are compared to those of the Groningen
group (Ref.[17], lines with circle labels). Displayed is the energy
difference to the potential
energy at momentum $k=0$ in each case.}
\label{fig:fig8}
\end{figure}


\begin{thebibliography}{99}
\bibitem{pb62} F.~Perey and B.~Buck, Nucl. Phys. {\bf 32} (1962) 353.
\bibitem{hod84} P.E.~Hodgson, Rep. Prog. Phys. {\bf 47} (1984) 613.
\bibitem{ham90} S.~Hama, B.C.~Clark, E.D.~Cooper, H.S.~Sherif, and
R.L.~Mercer, Phys. Rev. {\bf C 41} (1990) 2737.
\bibitem{ray92} L.~Ray, G.W.~Hoffmann, and W.R.~Coker, Phys. Rep. {\bf
212} (1992) 223.
\bibitem{mah91} C.~Mahaux and R.~Sartor, Adv. in Nucl. Phys. {\bf 20}
(1991) 1.
\bibitem{ger85} H.V.v.~Geramb (Ed): ``Medium Energy Nucleon and
Antinucleon Scattering'' (Springer, Berlin 1985).
\bibitem{sin75} B.~Sinha, Phys. Rep. {\bf 20} (1975) 1.
\bibitem{mah85} C.~Mahaux, P.F.~Bortignon, R.A.~Broglia, and
C.H.~Dasso, Phys. Rep. {\bf 120} (1985) 1.
\bibitem{bor92} M.~Borromeo, D.~Bonatsos, H.~M\"uther, and A.~Polls,
Nucl. Phys. {\bf A539} (1992) 189.
\bibitem{tor90} W.~Tornow, Z.P.~Chen, and J.P.~Delaroche, Phys. Rev.
{\bf C 42} (1990) 693.
\bibitem{mas91} C.~Mahaux and R.~Sartor, Nucl. Phys. {\bf A528}
(1991) 253.
\bibitem{ps64} F.~Perey and D.S.~Saxon, Phys. Let.. {\bf 10}
(1964) 107.
\bibitem{serot} B.D.~Serot and J.D.~Walecka,  Adv. in Nucl. Phys.
{\bf 16} (1986) 1.
\bibitem{rupr} R.~Machleidt, Adv. in Nucl. Phys. {\bf 19} (1989) 189.
\bibitem{shakin} M.R.~Anastasio, L.S.~Celenza, W.S.~Pong, and
C.M.~Shakin,  Phys. Rep. {\bf 100} (1978) 327; L.S.~Celenza and
C.M.~Shakin, ``Relativistic Nuclear Physics'' (World Scientific,
Singapore, 1986).
\bibitem{BM84} R.~Brockmann and R. Machleidt, Phys. Lett. {\bf 149B}
(1984) 283.
\bibitem{malf} B.~ter Haar and R.~Malfliet, Phys. Rep. {\bf 149}
(1987) 207.
\bibitem{jam80} M.~Jaminon, C.~Mahaux, and P.~Rochus, Phys. Rev. {\bf C
22} (1980) 2027.
\bibitem{fri1} R,~Fritz, H.~M\"uther, and R.~Machleidt, Phys. Rev.
Lett. {\bf 71} (1993) 46.
\bibitem{fri2} R.~Fritz and H.~M\"uther, Phys. Rev. {\bf C} in print.
\bibitem{ri83} L.~Rikus and H.V.~von Geramb, Nucl. Phys. {\bf A426}
 (1984) 496.
\bibitem{co90} W.R.~Coker and L.~Ray, Phys. Rev. {\bf C 42}
(1990) 659.
\bibitem{ar90} H.F.~Arellano, F.A.~Brieva, and G.~Love, Phys. Rev.
{\bf C 42} (1990) 652.
\bibitem{el90} Ch.~Elster, T.~Cheon, E.F.~Redish, and P.C.~Tandy, Phys.
Rev. {\bf C 41} (1990) 814.
\bibitem{ho85} C.J.~Horowitz, Phys. Rev. {\bf C 31} (1985) 1340.
\bibitem{co90b} W.R.~Coker and L.~Ray, Phys. Rev. {\bf C 42}
(1990) 2242.
\bibitem{ot91} N.A.~Ottenstein, E.F.~van Faassen, J.A.~Tjon, and
S.J.~Wallace, Phys. Rev. {\bf C 43} (1991) 2393.
\bibitem{li93} G.Q.~Li, R.~Machleidt,~R. Fritz, H.~M\"uther, and
Y.Z.~Zhuo, Phys. Rev. {\bf C 48} (1993) 2443.
\bibitem{m3y} G.~Bertsch, J.~Borysowicz, H.~McManus, and W.G.~Love,
Nucl. Phys. {\bf A 284} (1977) 399.
\bibitem{reid} R.~Reid, Ann. of Phys. {\bf 50} (1968) 411.
\bibitem{ellio} J.P.~Elliot, A.D.~Jackson, H.A.~Mavromatis,
E.A.~Sanderson, and B.~Singh, Nucl. Phys. {\bf A 121} (1968) 241.
\bibitem{czers}  P.~Czerski, H.~M\"uther, and W.H.~Dickhoff, J. Phys.
G, in print.
\bibitem{khoa} D.T.~Khoa and W.~von Oertzen, Phys. Lett. {\bf B 304}
(1993) 8.
\bibitem{nege} J.W.~Negele and D.~Vautherin, Phys. Rev. {\bf C 5}
(1972) 1472.
\bibitem{abramo} M.~Abramowitz and I.A.~Stegun, {\it Handbook of
Mathematical Functions} (Dover Pub. New York, 1972)
\bibitem{jam89} M.~Jaminon and C.~Mahaux, Rev. {\bf C 40} (1989) 354.
\bibitem{thesis} R.~Fritz, {\it Thesis} (University of T\"ubingen, in
preparation)
\bibitem{boro92} M.~Borromeo, D.~Bonatsos, H.~M\"uther, and A.~Polls,
Nucl. Phys. {\bf A 539} (1992) 189.
\bibitem{gua82} R.~Guardiola and J.~Ros, J. Comp. Phys. {\bf 45}
(1982) 390.
\bibitem{rolf} R.~Brockmann and H.~Toki, Phys. Rev. Lett. {\bf 68}
(1992) 340.
\bibitem{elsen} H.~Elsenhans, H.~M\"uther, and R,~Machleidt, Nucl. Phys.
{\bf A 515} (1990) 715.
\bibitem{mal94} H.F.~Boersma and R.~Malfliet, Phys. Rev. {\bf C 49}
(1994) 233.

\end{thebibliography}
\end{document}